\documentclass[10pt,english,british,tightenlines,eqsecnum,
floats,aps,amsmath,amssymb,nofootinbib,superscriptaddress,prd,showpacs,showkeys]
              {revtex4}
\usepackage[latin9]{inputenc}
\usepackage{color}
\usepackage{babel}
\usepackage{amsmath}
\usepackage{amssymb}
\usepackage{graphicx}
\usepackage{esint}
\usepackage[unicode=true,pdfusetitle,
 bookmarks=true,bookmarksnumbered=false,bookmarksopen=false,
 breaklinks=false,pdfborder={0 0 1},backref=false,colorlinks=false]
 {hyperref}

\makeatletter
\@ifundefined{textcolor}{}
{%
 \definecolor{BLACK}{gray}{0}
 \definecolor{WHITE}{gray}{1}
 \definecolor{RED}{rgb}{1,0,0}
 \definecolor{GREEN}{rgb}{0,1,0}
 \definecolor{BLUE}{rgb}{0,0,1}
 \definecolor{CYAN}{cmyk}{1,0,0,0}
 \definecolor{MAGENTA}{cmyk}{0,1,0,0}
 \definecolor{YELLOW}{cmyk}{0,0,1,0}
}

\usepackage{babel}

\usepackage{euscript}\usepackage{epsfig}

\usepackage{amsfonts}

\usepackage{fancyhdr}

\makeatother

\begin{document}
\title{\bf Neutrino mixing matrix and masses at a particular point of
the generalized Fridberg-Lee model }

\author{N. Razzaghi}
\email{n.razzaghi@qiau.ac.ir}

\affiliation{Department of Physics, Qazvin Branch, Islamic Azad
University, Qazvin, Iran}

\begin{abstract}
We propose the generalized Friedberg-Lee model neutrino mass model
at a particular point (point D at which
$\alpha=\beta=-\frac{1}{3}$) where at this point, the generalized
Fridberg-Lee model is converted to the Democratic mass matrix with
the $S_3$ symmetry. The Democratic texture has an experimentally
unfavored degenerate mass spectrum on the base of the tribimaximal
mixing matrix $(U_{TBM})$. We modify the Democratic mass matrix,
at point D, to obtain a nondegenerate mass spectrum by adding the
breaking mass term as preserving the twisted Fridberg-Lee
symmetry. Although the mixing matrix is still $(U_{TBM})$, where
leads to $\theta_{13}=0$ which is not consistent with the results
from Daya Bay and RENO experiments that have established a nonzero
value for $\theta_{13}$.

preserving the leading behavior of $U$ as tribimaximal, and we
apply the Broken Democratic neutrino mass texture as a mass matrix
at point D. Subsequently, we characterize a minimal perturbation
mass matrix which is responsible for a nonzero $\theta_{13}$ along
with CP violation parameters, besides the solar neutrino mass
splitting has been resulted from it. Let us mention that, unlike
other investigations, the perturbation matrix is not adopted on an
ad hoc basis, but is generated only in one step by the rules of
perturbation method that we will describe. Subsequently, we
develop the following results to the literature: (a) we obtain the
corresponding neutrino mixing matrix of the generalized
Fridberg-Lee model at point D with $\theta_{23}=\frac{\pi}{4}$ and
non-zero $\delta$; (b) the ordering of the neutrino masses is
inverted; (c) we also obtain the allowed range of the mass
parameters, the Dirac phase and the Jarlskog parameter which are
consistence with the available experimental data.

\end{abstract}


\date{Aug 14, 2023}


\maketitle

\section{Introduction}\label{sec1}
The results of the neutrino oscillation experiments \cite{exp,
exp1} have found that neutrinos have masses. The present
remarkable achievements $3\sigma$ global fits that led to the
existing and known neutrino oscillation parameters can be
summarized as follows\cite{exp}:

\begin{eqnarray}\label{exp}
\delta m^{2}[10^{-5}eV^{2}]&=&(6.94-8.14),\nonumber\\
|\Delta m^{2}|[10^{-3}eV^{2}]&=&(2.47-2.63)-(2.37-2.53),\nonumber\\
\sin^{2}\theta_{12}&=&(0.271-0.369),\nonumber\\
\sin^{2}\theta_{23}&=&(0.434-0.610)-(0.433-0.608),\nonumber\\
\sin^{2}\theta_{13}&=&(0.02000-0.02405)-(0.02018-0.02424),\nonumber\\
\delta&=&(128^\circ-359^\circ)-(200^\circ-353^\circ),
\end{eqnarray}

multiple sets of allowed ranges are stated, and the left columns
corresponds to normal hierarchy and the right columns to inverted
hierarchy. $\delta m^2\equiv m_2^2-m_1^2$ and $\Delta m^2\equiv
m_3^2-m_1^2 $.

The measurement indicate that $\theta_{13}$ is non-zero by more
than $5\sigma$ \cite{exp1} and is small compared to the other
neutrino mixing angles.

The lepton mixing matrix in the standard parametrization is \cite{mixing}:
\begin{equation}\label{emixing}
U_{PMNS}=\left(\begin{array}{ccc}c_{12}c_{13} & s_{12}c_{13} & s_{13}e^{-i\delta}\\
-s_{12}c_{23}-c_{12}s_{23}s_{13}e^{i\delta} &
c_{12}c_{23}-s_{12}s_{23}s_{13}e^{i\delta} &
s_{23}c_{13}\\s_{12}s_{23}-c_{12}c_{23}s_{13}e^{i\delta}
& -c_{12}s_{23}-s_{12}c_{23}s_{13}e^{i\delta} & c_{23}c_{13}\end{array}\right)\left(\begin{array}{ccc}e^{i\rho}  & 0 & 0 \\
0 & 1& 0\\0 & 0 & e^{i\sigma}\end{array}\right),
\end{equation}
where $c_{ij}\equiv\cos\theta_{ij}\text{ and
}s_{ij}\equiv\sin\theta_{ij}$ for $i,j=1,2,3 \hspace{2mm} (i<j)$;
$ \delta $ is known as the Dirac phase, analogous to the CKM
phase, and $\rho$ and $ \sigma$ are known as the Majorana phases
and are applicable to Majorana neutrinos.

A Dirac mass term for the neutrinos and charged leptons is written
as
 \begin{equation}\label{e2}\vspace{.2cm}
  {\cal L_\mathbf{m}}= {-\bar{\ell}_{{Li}}}{\cal
M_\mathbf{e}}^{ij}\ell_{Rj}-{\bar{\nu}_{{Li}}}{\cal
M_\mathbf{D}}^{ij}\nu_{Rj}+h.c.,
\end{equation}

Friedberge and Lee (FL) proposed a successful phenomenological
model of neutrino mass \cite{FL} with a suitable flavor symmetry
for Dirac neutrinos. In this model the charged-lepton mass matrix
is diagonal. Therefore, neutrino mixing matrix can simply be
described by a $ 3\times3 $ unitary matrix $U$ that transforms the
neutrino mass eigenstates into flavor eigenstates, $
(\nu_{e},\nu_{\mu},\nu_{\tau}).$
In the pure FL model, one of the neutrino masses is exactly zero,
which is partly responsible for the smallness of the neutrino
masses. Furthermore, assuming $\mu-\tau$ symmetry, the matrix $U$
reduces to the tribimaximal mixing matrix \cite{TBM}.

The Dirac neutrino mass operator of the FL model can be written as
\begin{eqnarray}\label{e3}\vspace{.5cm}
  {\cal M_\mathbf{FL}}&=&a\left(\bar{\nu}_{\tau}-\bar{\nu}_{\mu}\right)\left(\nu_{\tau}-\nu_{\mu}\right)
+ b\left(\bar{\nu}_{\mu}-\bar{\nu}_{e}\right)\left(\nu_{\mu}-\nu_{e}\right)\nonumber\\
&+&
c\left(\bar{\nu}_{e}-\bar{\nu}_{\tau}\right)\left(\nu_{e}-\nu_{\tau}\right)
+
m_{0}\left(\bar{\nu}_{e}\nu_{e}+\bar{\nu}_{\mu}\nu_{\mu}+\bar{\nu}_{\tau}\nu_{\tau}\right).
\end{eqnarray}
All the parameters in this model ($a,b,c$ and $m_{0}$) are assume
to be real. For $m_{0}=0$, this Lagrangian has the following
symmetry $
 \nu_{e}\rightarrow\nu_{e}+z $, $\nu_{\mu}\rightarrow\nu_{\mu}+z $, and $  \nu_{\tau}\rightarrow\nu_{\tau}+z $,
 where $z$ is an element of the Grassman algebra. In a special case where $z={\rm constant}$, we get a FL symmetry \cite{FL}, whose kinetic term is also invariant.
  However the other terms of the electroweak Lagrangian do not
 have such symmetry. The $m_{0}$ term explicitly breaks this symmetry.

 However, we should mention that the FL symmetry leads into a magic matrix and this property is not broken by the $m_{0}$ term.
 The magic symmetry has many demonstrations which
 we will talk about in details. It has also been argued that the FL symmetry is the residual symmetry of the neutrino mass
 matrix after the $SO(3)\times U(1)$ flavor symmetry breaking \cite{FL2}. The mass matrix can be displayed by,
 \begin{equation}\label{e4}
M_{FL} =\left(\begin{array}{ccc}b+c+m_{0} & -b & -c\\
-b & a+b+m_{0} & -a\\-c & -a & a+c+m_{0}\end{array}\right),
\end{equation}
where $a \propto\left(Y_{\mu\tau}+Y_{\tau\mu }\right)$, $b
\propto\left(Y_{e\mu}+Y_{\mu e }\right)$ and $  \ c
\propto\left(Y_{\tau e}+Y_{e\tau }\right) $ and $Y_{\alpha\beta}$
denotes the Yukawa coupling constant \footnote{The proportionality
constant is the expectation value of the Higgs field.}. It is
obvious that $ M_{FL} $ has an exact $\mu-\tau$ symmetry only if
$b=c$. Setting $b=c$ and using the hermiticity of $M_{FL}$, a
straightforward diagonalization procedure yields
$U^{T}_{TBM}M_{FL}U_{TBM}=\text{ Diag }\{m_{1},m_{2},m_{3}\} $,
where
\begin{equation}\label{e6}\vspace{.2cm}
m_{1}=3b+m_{0}~~~~m_{2}=m_{0}~~~~m_{3}=2a+b+m_{0}.
\end{equation}
Therefore, the well-known tribimaximal (TBM) neutrino mixing
matrix can be reproduced \cite{TBM}, which can be written as
\begin{equation}\label{etbm}
U_{TBM} =\left(\begin{array}{ccc}-\sqrt{\frac{2}{3}} & \frac{1}{\sqrt{3}} & 0\\
\frac{1}{\sqrt{6}} & \frac{1}{\sqrt{3}} &
-\frac{1}{\sqrt{2}}\\\frac{1}{\sqrt{6}} & \frac{1}{\sqrt{3}} &
\frac{1}{\sqrt{2}}\end{array}\right).
\end{equation}

The exact tribimaximal mixing matrix $U_{TBM}$ sets
$(U_{e3})_{TBM}=0$ independently of the model. The values of the
mixing angles in the mixing matrix $U_{TBM}$ agree with the
current existing neutrino oscillation parameters except $U_{e3}$.
The role of a non-zero $\theta_{13}$ or equivalently $U_{e3}$, is
numerus. It is necessary for CP violation in neutrino oscillations
and can explain leptogenesis. Of course, for CP violation, both
$\theta_{13}$ and the complex phase $\delta$ should be non-zero.
Moreover, $\theta_{13}\neq0$ corresponds to the quark sector since
the mixing between the three generations is a confirmed result,
although the mixing angles in the lepton sectors are very small in
comparison to the quark sector. It is obvious that
Eq.\,(\ref{etbm}) implies the following forms of the neutrino mass
matrix in the flavor basis,
\begin{equation}\label{em3}
{\cal{M}} =\frac{m_1}{6}\left(\begin{array}{ccc}4 & -2 & -2\\
-2 & 1 & 1\\-2 & 1 & 1\end{array}\right)+\frac{m_2}{3}\left(\begin{array}{ccc}1 & 1 & 1\\
1 & 1 & 1\\1 & 1 & 1\end{array}\right)+\frac{m_3}{2}\left(\begin{array}{ccc}0 & 0 & 0\\
0 & 1 & -1\\0 & -1 & 1\end{array}\right),
\end{equation}
where $m_i~(i=1,2,3)$ are the neutrino mass eigenvalues. The
second term in Eq.\,(\ref{em3}) is noteworthy. This form of
neutrino mass matrix is called the democratic matrix \cite{demo1}.
Democratic matrix is a phenomenological model of Dirac neutrino
mass with $S(3)_L\times S(3)_R$ flavor symmetry. A democratic
basis is adopted to produce a flavor--democratic mass matrix in
which all elements of the matrix are equal. In this basis, the
$S_3$ operations generate the permutations of three objects, e.g.,
the exchanging the first and second indices. Invariance under such
transformations requires the universal size of the
three-generation couplings when belonging to a three-dimensional
representation of $S_3$ and the Higgs field is in the singlet.

The smallness of $\theta_{13}$ compered to the other mixing angles
persuade us to modify the neutrino mixing matrix by a small
perturbation in the basic tribimaximal structure and could lead to
a realistic neutrino mixing matrix. We focus our interest on the
neutrino mass model proposed by Friedberg and Lee (FL) in a very
special case, which leads us to obtain the Democratic mass matrix
in the flavor basis. The democratic matrix is suitable for
generating $\theta_{13}\neq0$ from an initial tribimaximal form.
However, the result of its mass spectrum with the $U_{TBM}$ is
experimentally unfavorable at first glance. Therefore, the main
objective of this paper is to find the experimentally favored mass
spectrum of the Democratic matrix together with the realistic
neutrino mixing matrix through a small perturbation in the basic
tribimaximal framework. There are many neutrino mass models from
which the tribimaximal form of the neutrino mixing matrix \cite{8}
can be obtained. Moreover, many theoretical and phenomenological
works have discussed massive neutrino models to generate
$\theta_{13}\neq0$ in different ways starting from an initial
tribimaximal form \cite{ttbm}. To have CP-violation in the
standard parametrization given in Eq.\,(\ref{emixing}), the
necessary conditions are $\delta\neq0$ and $\theta_{13}\neq0$.

The Jarlskog rephasing invariant parameter $J$ \cite{J},

\begin{equation}\label{eJ}\vspace{.2cm}
J={\cal{I}}m(U_{11}U^{\ast}_{12}U^{\ast}_{21}U_{22}),
\end{equation}

is relevant for CP violation in lepton number conservation
processes like neutrino oscillations.
Oscillation experiments cannot distinguish between the Dirac and
Majorana neutrinos \footnote{ Detection of neutrinoless double
beta decay would provide direct evidence for lepton number
non-conservation and the Majorana nature of neutrinos.}. Numerous
theoretical and phenomenological works have discussed  massive
neutrino models breaking $\mu-\tau$ symmetry as a prelude to CP
violation\cite{theoretical}.

In this paper, we generalize the FL model by introducing complex
parameters. We work on the massive FL model and set some obvious
constraints that mass eigenvalues become real. In generalized FL
model, there is a confined region of the parameter space where CP
violation arises. However, we focus on a notable point, point D,
on the border of this region that the neutrino mass matrix is
changed to the Democratic neutrino mass matrix in the parameter
space. The democratic neutrino mass matrix is not consistent with
the experimental data $m_1=m_3=0$. Therefore, we consider a case
with $m_1\neq 0$ by adding the symmetry breaking term based on FL
symmetry. Consequently, the universality of the experimental data
imposes some constraints on the elements of the mass matrix in the
basic structure of $U_{TBM}$. Subsequently, we obtain our
unperturbed neutrino mass matrix with $\mu-\tau$ symmetry and
magic property in the flavor basis. However, we find that the
solar mass splitting is absent, hence we generate this splitting
by a small perturbation, which is responsible for $U_{e3}\neq0$.
To do this, we produce the perturbation mass matrix by applying
the perturbation theory in the mass basis with complex elements,
which breaks mildly the $\mu-\tau$ symmetry and this may generate
$U_{13}\neq0$ along with CP violation paremeter. So we obtain
nonzero values for both $\theta_{13}$ and $\delta$ which, together
with a realistic neutrino mixing matrix, leads to a CP violation.

The outline of the paper is as follows. In section 2, we present
our model and display the condition of a considerable point in the
generalized FL model, and then present the results of complex
perturbation analysis described above. Moreover, we obtain the
perturbation mass matrix generating CP violation, solar mass
splitting and we get a realistic neutrino mixing matrix.
In section 3, we display that our model at point D is, in general,
consistent with the available experimental data. Afterwards, we
find the allowable ranges for all the parameters of the model.
Ultimately, not only do we check the consistency of all of the
results with the available experimental data, but also present our
predictions for the actual masses, the Dirac phase, and the
Jarlskog parameter. In section 4, we summarize and present the
results. In Appendix A, we briefly introduce Twisted Friedberg-Lee
symmetry.

\section{The Model}\label{sec2}
In this section, we generalize the FL model by adding complex
Yukawa coupling constants to procure CP violation, which is
achieved by obtaining $ U_{e3}\neq0 $. However, we suppose that
the eigenvalues of the generalized mass matrix are real. We figure
out that only one specific choice allows for the softly breaking
of $\mu-\tau$ symmetry, {\em i.e.} $(a\in\Re;~ b, c\in\mathbb{C}$
and $b=c^{\star})$ which leads to a non-hermitian mass matrix. To
simplify the notation, we define the parameters as follows:
$\Re\left(b\right)=\Re\left(c\right)= b_{r} \text{ and
}\Im\left(b\right)=-\Im\left(c\right)= B$\footnote{ The parameters
that distinguish the measure of CP violation and $\mu-\tau$
symmetry breaking are proportional to $B$ and so we expect them to
be small.}.

The neutrino mass matrix $ M^{\prime}_{\nu}$ is given by
\begin{equation}\label{e8}
M^{\prime}_{\nu}=\left(\begin{array}{ccc}2b_{r}+m_{0} & -b_{r} & -b_{r}\\
-b_{r} & a+b_{r}+m_{0} & -a\\-b_{r}
& -a & a+b_{r}+m_{0}\end{array}\right)+iB\left(\begin{array}{ccc}0 & -1 & 1\\
-1 & 1 & 0\\1 & 0 & -1\end{array}\right).
\end{equation}
Note that $M^{\prime}_{\nu}$ is a symmetric matrix,therefore it
could also be used as a Majorana mass matrix. In addition
$M^{\prime}_{\nu}$ and $M_{FL}$ are magic matrices since one of
the eigenstates is $(\frac{1}{\sqrt{3}}, \frac{1}{\sqrt{3}},
\frac{1}{\sqrt{3}})$. We choose it to be $|\nu_{2}\rangle$ to be
consistent with Eq.\,(\ref{etbm}).

A naive diagonalization of $M^{\prime}_{\nu}$ yields,
\begin{eqnarray} \vspace{.2cm}\label{e11}
\breve{m}_{1}&=&(a+2b_{r}+m_{0})+\sqrt{(a-b_{r})^{2}-3B^{2}}\nonumber\\\breve{m}_{2}&=&m_{0},
\nonumber\\\breve{m}_{3}&=&(a+2b_{r}+m_{0})-\sqrt{(a-b_{r})^{2}-3B^{2}}.
\end{eqnarray}

The results given in Eq.\,(\ref{e11}) are correct only in the
limit $B\rightarrow0$,
when we compare the results in this limit with those in
Eq.\,(\ref{e6}), we find that $ a<b $.

Since $M^{\prime}_{\nu}$ is a non-hermitian matrix, we require two
distinguished unitary matrices $U$ and $V$ to diagonalize
it.\footnote{ These matrices can be obtained by diagonalizing $
M^{\prime}_{\nu}{M^{\prime}_{\nu}}^{\dag} $ and $
{M^{\prime}_{\nu}}^{\dag}M^{\prime}_{\nu} $, separately. $U$ and
$V$ are the conventional transformation matrices for left-handed
and right-handed neutrinos, respectively.} that $V = U^{\ast}$.
The resulting valid diagonal matrix is acquired by
$M^{\prime}_{diag}=U^{\dag}M^{\prime}_{\nu}V$ and its elements
are:
\begin{eqnarray} \vspace{.2cm}\label{e111}
m'_{1}&=&\frac{iB(a-b_{r})+3B^{2}+(a+2b_{r}+m_{0})^{2}-(a-b_{r}+iB)
\sqrt{3B^{2}+(a+2b_{r}+m_{0})^{2}}}{a+2(b_{r}+iB)+m_{0}},
\nonumber\\m'_{2}&=&m_{0},\nonumber\\m'_{3}&=&\frac{iB(a-b_{r})+3B^{2}
+(a+2b_{r}+m_{0})^{2}+(a-b_{r}+iB)\sqrt{3B^{2}+(a+2b_{r}
+m_{0})^{2}}}{a+2(b_{r}+iB)+m_{0}},
\end{eqnarray}

where $m'_{1}$ and $m'_{3}$ are complex\footnote{ In the Dirac
case \cite{phase1}, we can extract the phases and transform them
into the mass eigenstates.}. The most general form of the diagonal
neutrino mass matrix can be written as,
\begin{equation}\label{ereal}
M^{\prime}_{diag}=e^{i\alpha}e^{i\beta\lambda_{3}}e^{i\gamma\lambda_{8}}M'^{real}_{diag}.
\end{equation}

\begin{center}
\begin{figure}[th] \includegraphics[width=6cm]{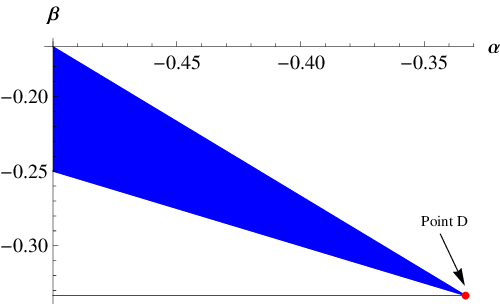}
\caption{\label{fig.1} \small
   CP violation is only possible in the right-angled triangle in the $\alpha-\beta$ parameter space, where $\alpha\equiv\frac{a}{m_{0}}$ and $\beta\equiv\frac{b_{r}}{m_{0}}$.
   The specified triangle shows the allowed region within our model. The red point shows our remarkable point D at $(-\frac{1}{3},-\frac{1}{3})$. (The line above the base of the triangle is given by $2b_{r}+a+m_{0}=0$).}
  \label{geometry}
\end{figure}
\end{center}

$\alpha$ is an overall phase, in our model it automatically turns
out to be zero.
We would not use the overall phase even if it were not zero. Due
to the fact that $m'_{2}$ is real, we acquire $\beta=\gamma$. This
leads that the $\arg(m'_{1})=-\arg(m'_{3})=2\beta$\footnote{
Note that ignoring the overall phase is equivalent to the
following: $Det(M^{\prime}_{diag})$ is real and $Det
(U)=1$\cite{phase1}.}. Using this condition in Eq.(11) we get
\begin{equation}\label{e12}
B=\pm\sqrt{\frac{-\left(2a+b_{r}+m_{0}\right)\left(3b_{r}+m_{0}\right)}{3}}
\end{equation}
To require that the quantity $B$ take on real values, one must
constrain $b_{r}$ as $-\frac{m_{0}}{3}\leq b_{r} \leq -(2a+m_{0})
$, where the lower bound of $b_{r}$ is a check on the condition
$m_{1}>0$ in Eq.\,(\ref{e6}). Since $U$ and $V$ should approach
$U_{TBM}$ (given by Eq.\,(\ref{etbm})) in the limit
$B\rightarrow0$, we get $2b_{r} + a +m_{0}\geq0$. Using
\eqref{e12} and $ a<b $, we get $3a+m_{0}\leq0$. Finally, from the
allowed regions for the parameters of the model and considering
the overall symmetry of the F.L model we get $b_r<0$\footnote{
This conclusion is consistent with the results of the experiments
on the oscillation of solar neutrinos, which indicate that
$m_{2}>m_{1}$.}. We have to take into account that a CP violation
can only occur in a restricted range in the plane $a$-$b_{r}$ with
$B\neq0$ . In the figure (\ref{fig.1}) we have shown the allowed
region of the parameter space in which the CP violation occurs.

A detailed study of this region, its corresponding CP violation,
achievable consequences, and its agreement with experimental data
has been presented in Ref. \cite{me1}.

In Figure (\ref{fig.1}) there is a remarkable point (red point)
which displays
 $ a=b_{r}=-\frac{m_{0}}{3} $ where at it $B=0$.  Therefore at
 first glance, it seems we cannot have CP violation at it. We have called this notable Point, point D.

The study of point D is significant because at this point we have
$\Delta m^2=0$.

In the following, we focus our study only at point D (see Figure
(\ref{fig.1}))  where the generalized FL neutrino mass matrix,
$M^{\prime}_{\nu}$ in Eq.\,(\ref{e8}), is reduced to the following
specific form:
\begin{eqnarray}\label{e22}
{M_{\nu}}|_{\text{point
D}}&=&(3b_{r}+m_{0})\left(\begin{array}{ccc}1 & 0 & 0\\
0 & 1 & 0\\0 & 0 & 1\end{array}\right)-b_{r}\left(\begin{array}{ccc}1 & 1 & 1\\
1 & 1 & 1\\1 & 1 &
1\end{array}\right)\nonumber\\&=&(3b_{r}+m_{0})\textbf{1}-b_{r}\mathbf{D}.
\end{eqnarray}
The first term in Eq.\,(\ref{e22}) certainly vanishes and
$M_{\nu}$ reduces to $-b_{r}\mathbf{D}$, $\mathbf{D}$ denotes
democratic mass matrix.
Therefore $M_{\nu}$ obtains the democratic form that can be
realized by the $S_3$ family of symmetric permutations
\cite{demo1}. Here we separately impose the $S_3$ permutation
symmetry for the left- and right- handed neutrinos as flavor
symmetry, which are indicated by $S_{3L}$ and $S_{3R}$,
respectively. Therefore it leads to $M_{\nu}$ in Eq.\,(\ref{e22})
invariant under $S_{3L}\times S_{3R}$ symmetry.

$M_{\nu}$ in Eq.\,(\ref{e22}) can be diagonalized by $U_{TBM}$
that yields $m_2=m_0$ and $m_1=m_3=0$. Since such a neutrino mass
matrix is experimentally unfavorable, we should rescue $M_{\nu}$
by replacing it with an alternative one (in which  $m_1$ or $m_3$
does not vanish) by using a possible procedure. With that in mind,
let us consider a non-zero $m_1$ case below by adding the
symmetry-breaking term based on FL symmetry.


The combination of FL symmetry with $\mu-\tau$ symmetry is a kind
of familiar translational symmetry called twisted FL
symmetry\cite{TFL}. The twisted FL symmetry for Dirac neutrinos is
separately imposed on left-handed and right-handed neutrinos as
follows:
\begin{eqnarray}\label{eq:RFL}
&&\nu_{Li} \rightarrow \nu_{Li}^{'}=S_{ij}^L\nu_{Lj}+\Lambda_{Lj} z\ \nonumber\\
&&\nu_{Ri} \rightarrow \nu_{Ri}^{'}=S_{ij}^R\nu_{Rj}+\Lambda_{Rj}
z\ ,
\end{eqnarray}
where $\Lambda=(\Lambda_1,~ \Lambda_2,~ \Lambda_3)^T$ are
c-numbers, $z$ is a space-time independent Grassmann parameter
such that $z^2=0$, and $S$ is the permutation matrix between the
second and third families, as follows
\begin{eqnarray}
 S=\left(\begin{array}{ccc}
  1 & 0 & 0 \\
  0 & 0 & 1 \\
  0 & 1 & 0
 \end{array}\right). \label{eq:s}
\end{eqnarray}
Therefore, As mentioned in Appendix A under these transformations,
the breaking mass matrix for the $S_{3L}\times S_{3R}$ flavor
symmetry of $M_{\nu}$ in Eq.\,(\ref{e22}) obtain as follows
\begin{eqnarray}
 M^B_{\nu}=g\left(\begin{array}{ccc}
  4 & -2 & -2 \\
  -2 & 1 & 1 \\
  -2 & 1 & 1
 \end{array}\right). \label{eq:eFL4}
\end{eqnarray}

 We added the breaking mass matrix, $M^B_\nu$ in Eq.\,(\ref{eq:eFL4}), to $M_\nu$ in Eq.\,(\ref{e22}) as a breaking
 mass matrix for the $S_{3L}\times S_{3R}$ flavor symmetry. Consequently, the
total neutrino mass matrix at point D is written as follows
\begin{eqnarray}
 {M_\nu^{BD}}|_{\text{point
D}}=\frac{m_0}{3}
 \left(\begin{array}{ccc}
      1 & 1 & 1 \\
      1 & 1 & 1 \\
      1 & 1 & 1
  \end{array}\right)+g
 \left(\begin{array}{ccc}
     4 & -2 & -2 \\
     -2 & 1 & 1 \\
     -2 & 1 & 1
  \end{array}\right)\ .\label{eq:MD}
\end{eqnarray}
We call it Broken Democtratic (BD) neutrino mass matrix and its
allowable neutrino mass spectrum is given by
\begin{eqnarray}
\tilde{ m}_1 = 6g,\ \ \tilde{m}_2 =m_0,\ \ \tilde{m}_3 = 0\
.\label{eq:MD1}
\end{eqnarray}

Note that, up to now, the achievements are: (i) the twisted FL
symmetry is employed as a breaking mass matrix for the
$S_{3L}\times S_{3R}$ flavor symmetry, (ii) the total mass matrix
in Eq. (\ref{eq:MD}) violates the $S_{3L}\times S_{3R}$ symmetry
but preserves the $\mu-\tau$ symmetry and magic property.
Therefore, still $U_{TBM}$ can be realized as the mixing matrix
and (iii) the neutrino masses ordering is inverted.

Thus, the main objective will be to find a small perturbation mass
matrix together along with a realistic neutrino mixing matrix that
is corresponding $ M_\nu^{BD}$ in Eq. (\ref{eq:MD}) on the basic
tribimaximal mixing matrix. Therefore, according to
Eq.\,(\ref{em3}), a general mass matrix $\cal{M}$ satisfies
$U_{TBM}$ when represented in the flavor basis and its most
general form is,
\begin{eqnarray}\label{egf0}
\cal{M}&=&U_{TBM}\left(\begin{array}{ccc}m_1 & 0 & 0\\
0 & m_2 & 0\\0 &0& m_3\end{array}\right)U^T_{TBM}\nonumber
 \nonumber
\\&=&\left(\begin{array}{ccc}A & B & B\\
 B& A+C &  B-C\\B
 &  B-C & A+C\end{array}\right),
\end{eqnarray}
where
\begin{eqnarray}\label{eABC}
A&=&m-\frac{\Delta_{31}}{3},\nonumber
\\B&=&\frac{\Delta_{31}-\Delta_{32}}{3},~~~~\text{and}~~~~C=\frac{\Delta_{31}}{2}.
\end{eqnarray}
and we have set
\begin{eqnarray}\label{egf}
m &=&\frac{\sum m_i}{3},\nonumber
\\\Delta_{3j}&\equiv&(m_3-m_j),~~~~\text{for}~~~~j=1,2~.
\end{eqnarray}
Substituting the $\tilde{m}_{i}$s in Eq.\,(\ref{eq:MD1}) into
(\ref{egf}), we obtain $m=\frac{m_0}{3}+2g$, $\Delta_{32}\equiv
-m_0$ and $\Delta_{31}\equiv -6g$. We work in a flavor basis in
which mixing in the lepton sector is determined entirely by the
neutrino mass matrix.

Experimental data has definitely confirmed that $\delta
m^2=m^2_2-m^2_1>0$ and  it is too small. Substituting
$\tilde{m}_{i}$s in Eq.\,(\ref{eq:MD1}) into $\delta m^2$, we
obtain $|g|<\frac{m_0}{6}$. Then assuming
$\Delta_{32}\simeq\Delta_{31}\equiv\Delta$, we obtain
$\Delta\simeq-m_0\simeq-6g$, which indicates
$g\simeq\frac{m_0}{6}$, and $\Delta$ is negative\footnote{ We
should note that $\Delta$ is an important quantity due to the
scale for atmospheric neutrino oscillations is set by it.}. In the
following by using these approximations the neutrino mass matrix
$M_\nu^{BD} $ in Eq.\,(\ref{eq:MD}) in the flavor basis rewritten
as follows
\begin{equation}\label{eunp}
{M^{0}_{\nu}}|_{\text{point
D}} \simeq \left(\begin{array}{ccc}m-\frac{\Delta}{3} & 0 & 0\\
0 & m+\frac{\Delta}{6} & -\frac{\Delta}{2}\\0 &-\frac{\Delta}{2}&
m+\frac{\Delta}{6}
\end{array}\right),
\end{equation}
here in $M^{0}_{\nu}$ is the unperturbed neutrino mass matrix with
magic $\mu-\tau$ symmetry (at point D). Therefore, The mixing
matrix corresponding to $M^{0}_{\nu}$ is still $U_{TBM}$.

The eigenvalues of $M^{(0)}_{\nu}$ in Eq.\,(\ref{eunp}) are obtain
as follows,
\begin{eqnarray}\label{emm2}\vspace{.2cm}
m^{(0)}_1 \simeq m^{(0)}_2 =m-\frac{\Delta}{3}=m_0,
~~~~\text{and}~~~m^{(0)}_{3}=m+\frac{2\Delta}{3}=0.
\end{eqnarray}
 $m^{(0)}_1$, and $m^{(0)}_2$
are real and positive, although at this level, the solar mass
splitting is absent. Also $m_3=0$, similar to the result obtained
in \cite{me1}, therefore the ordering of neutrino masses is still
inverted.

 We should note that, up to now, the shortcomings are:
(i) the absence of the solar neutrino mass splitting, and (ii) the
mixing matrix is still $U_{TBM}$. Thus, we will aim to obtain the
solar mass splitting by the same mass perturbation that is the
cause of $\theta_{13} \neq 0$ and CP violation. Finally, in this
procedure CP violation conditions necessarily mandate that
$\mu-\tau$ symmetry should be broken. An interesting question is
whether $\theta_{23}=45^{\circ}$ holds after the $\mu-\tau$
symmetry breaking.

In summary, up to now, we have proposed that the generalized FL
neutrino mass matrix in Eq.\,(\ref{e8}) at point D (see Figure
(\ref{fig.1})) has a combination in which $m_1$, $m_3$, and CP
violation parameters are vanishing, while
$\theta_{23}=45^{\circ}$. Moreover, $\delta m^2$ does not vanish
but is not consistent with experimental data. However, the solar
mixing angle $\theta_{12}$ can be chosen by the mixing angles of
the tribimaximal mixing matrix. This is a promising estimate of
the observed data although some characteristics are missing at
point D. Therefore, at this notable point the neutrino mass matrix
in Eq.\,(\ref{e22}) has two mass eigenvalues, $m_1$ and $m_3$,
which are degenerate, hence it is highly distinctive from the
obtained results of the generalized FL model in \cite{me1} and the
experimental data \cite{exp}, in which $m_1\neq m_3$. In the
following, we obtain the allowable neutrino mass spectrum of the
generalized FL model at point D by adding the breaking mass matrix
for the $S_{3L}\times S_{3R}$ flavor symmetry in to the neutrino
mass matrix in Eq.\,(\ref{e22}). Then, according to the available
experimental data for the mass of neutrinos, we apply some
approximations to the elements of the mass matrix $M_\nu^{BD}$ in
Eq.\,(\ref{eq:MD}). Therefore, we obtain the mass matrix
$M_\nu^{0}$ in Eq.\,(\ref{eunp}) that its mass spectrum shows the
absence of the solar neutrino mass splitting. In the next stage,
we will consider obtaining a perturbation matrix, which could
generate not only CP violation parameters in the neutrino mixing
matrix, namely, $U_{13}$, ($\theta_{13}$ and $\delta$) but also
$\delta m^2$ which provides only minor corrections to the mixing
angle $\theta_{12}$ (but not to $\theta_{23}$). We believe,
because of the small $\theta_{13}$ and $\delta m^2$, the
perturbation method is a more meticulous method than others for
finding the correction of the tribimaximal mixing matrix.
Moreover, to the best of our knowledge, this kind of generalized
FL model at a specific point has not been performed in the
literature yet.


As mentioned before, our method for establishing the structure of
the neutrino mass matrix, at point D, is to apply perturbation
theory. Thus, we have ${M_\nu}|_{\text{point
D}}=M_\nu^{0}+M_\nu^{P}$ , where $M_\nu^{P}$ is a perturbation
matrix and $M_\nu^{P}<<M_\nu^{0}$. Both $M_\nu^{0}$ and
$M_\nu^{P}$ will be symmetric and could, in general, be complex.
However, in our case $M_\nu^{0}$ in Eq.\,(\ref{eunp}) is a
Hermitian matrix (real and symmetric). The following will consider
the case of complex $M_\nu^{P}$. Therefore, in this situation, we
have $\theta_{13} \neq 0$ and $\delta\neq0$ which means CP is
violated and the solar neutrino masses are split.

The eigenstates of $M_\nu^{0}$ in the mass basis, the unperturbed
mass eigenstates, are as follows,

\begin{equation} \vspace{.2cm}\label{em0}
|\nu^{(0)}_{1}\rangle=\left(\begin{array}{ccc}1\\
0\\0
\end{array}\right), ~~~~|\nu^{(0)}_{2}\rangle=\left(\begin{array}{ccc}0\\
1\\0\end{array}\right), ~~~~~|\nu^{(0)}_{3}\rangle=\left(\begin{array}{ccc}0 \\
0\\1\end{array}\right),
\end{equation}
in which $|\nu^{(0)}_{1}\rangle$, and $|\nu^{(0)}_{2}\rangle$  are
degenerate. We investigate $M_\nu^{P}$ such that the first two
mass eigenstates in Eq.\,(\ref{em0}) are its non-degenerate
eigenstates, namely,
$\langle\nu_{i}^{(0)}|M_{\nu}^{P}|\nu_{j}^{(0)}\rangle=
m_{i}^{(1)}\delta_{ij}$ where $ (i,j=1,2)$, with $m_1^{(1)}\neq
m_2^{(1)}$. Therefore, in the mass basis, we take
$(M_\nu^{P})_{33}=0$ and consequently consider only
$(M_\nu^{P})_{13}$ and $(M_\nu^{P})_{23}$. So the basis vectors
$|\nu^{(0)}_{1}\rangle$, and $|\nu^{(0)}_{2}\rangle$ are chosen
with the aim that they regenerate the correct solar mixing and the
physical basis is fixed by perturbation. Needless to say, when
eigenstates in Eq.\,(\ref{em0}) expressed in the flavor basis are
the columns of $U_{TBM}$ in Eq.\,(\ref{etbm}). Therefore in the
flavor basis, eigenstates are as follows,

\begin{equation} \vspace{.2cm}\label{em00}
|\nu^{(0)}_{1}\rangle=\left(\begin{array}{ccc}\sqrt{\frac{2}{3}}\\
-\frac{1}{\sqrt{6}}\\\frac{1}{\sqrt{6}}
\end{array}\right), ~~~~|\nu^{(0)}_{2}\rangle=\left(\begin{array}{ccc}\frac{1}{\sqrt{3}}\\
\frac{1}{\sqrt{3}}\\-\frac{1}{\sqrt{3}}\end{array}\right), ~~~~~|\nu^{(0)}_{3}\rangle=\left(\begin{array}{ccc}0 \\
\frac{1}{\sqrt{2}}\\\frac{1}{\sqrt{2}}\end{array}\right).
\end{equation}

our goal is to obtain the third perturbed mass eigenstate in the
CP violated case that when expressed in the flavor basis is the
third column of the mixing matrix in Eq.\,(\ref{emixing}). Thus we
can obtain the elements of the perturbation mass
matrix~\footnote{we do not set a perturbation mass matrix by hand;
instead, we obtain it by using the third perturbed mass
eigenstate}.

we suppose $M_\nu^{P}$ which is symmetric, to be a complex matrix,
and therefore not Hermitian, and this is true as well as for the
total mass matrix $M_\nu=M_\nu^{0}+M_\nu^{P}$. Therefore, this is
accomplished by obtaining nonzero values for $\theta_{13}$ and
$\delta$. The columns of the mixing matrix $U$ in the
Eq.\,(\ref{emixing}) must be eigenvectors of $M_\nu^\dagger M_\nu
= M_\nu^{0\dagger}M_\nu^0 + M_\nu^{0\dagger}M_\nu^P +
M_\nu^{P^{\dagger}}M_\nu^0$, where we have dropped the term ${\cal
O} (M_\nu^P)^2$. In the following, we should mention that
unperturbed $M_{\nu}^{0^{\dagger}}M_{\nu}^{0}$ is Hermitian and
its eigenstates are the same as the columns that produce
$U_{TBM}$, in the Eq.\,(\ref{etbm}), and its eigenvalues are
$|m_{1}^{(0)}|^{2}$, $|m_{2}^{(0)}|^{2}$, and $|m_{3}^{(0)}|^{2}$.
For the perturbation expansion we keep terms up to linear order in
$s_{13}$. To first order we have
\begin{equation}
|\nu_3\rangle=|\nu^{(0)}_3\rangle+\sum_{j \neq 3} {\cal
C}_{3j}|\nu^{(0)}_j\rangle \;. \label{eper1}
\end{equation}
Where,
\begin{equation}
{\cal C}_{3j}= -{\cal C}^*_{j3}=\left(|m^{(0)}_3|^2 -
|m^{(0)}_j|^2\right)^{-1}{\cal M}_{j3}, \;\; (j \neq 3) \;
 \label{eper3}
\end{equation}
where ${\cal M}_{j3}={<\nu^{(0)}_j| ( M_\nu^{0\dagger}M_\nu^P +
M_\nu^{P^{\dagger}}M_\nu^0) |\nu^{(0)}_3> }$ and the coefficients
${\cal C}_{3j}$ are complex and proportional to ${\cal M}_{j3}$ in
the mass basis \cite{meuni}.

Evidently, $|\nu_3\rangle$ in Eq.\,(\ref{eper1}) should correspond
to the third column of the neutrino mixing matrix in
Eq.\,(\ref{emixing}). we can quickly determine ${\cal C}_{31}$ and
${\cal C}_{32}$ by using Eq.\,(\ref{eper1}) in the flavor basis.
Therefore explicitly we obtain the matrix equation, as follows
\begin{equation}\label{e1}
\left(\begin{array}{ccc}s_{13}e^{-i\delta}\\
s_{23}c_{13}\\c_{23}c_{13}
\end{array}\right)=\left(\begin{array}{ccc}0\\
-\frac{1}{\sqrt{2}}\\\frac{1}{\sqrt{2}}\end{array}\right)+\left(\begin{array}{ccc}\frac{-\sqrt{2}{\cal C}_{31}+{\cal C}_{32}}{\sqrt{3}}\\
\frac{{\cal C}_{31}}{\sqrt{6}}+\frac{{\cal
C}_{32}}{\sqrt{3}}\\\frac{{\cal C}_{31}}{\sqrt{6}}+\frac{{\cal
C}_{32}}{\sqrt{3}}\end{array}\right)
\end{equation}
Readily one obtains, to order linear in $s_{13}$, ${\cal
C}_{31}=-\sqrt{\frac{2}{3}} s_{13}e^{-i\delta}$ and ${\cal
C}_{32}=\sqrt{\frac{1}{3}} s_{13}e^{-i\delta}$, where maximality
of the 2-3 mixing angle has been used ($\theta_{23}=45^\circ$).
Therefore, in the mass basis by using Eq.\,(\ref{emm2}) and
Eq.\,(\ref{eper3}) we obtain,
${(M^{P}_\nu)}_{13}=~m_0\sqrt{\frac{2}{3}} s_{13}e^{-i\delta}$ and
${(M^{P}_\nu)}_{23}=~-m_0\sqrt{\frac{1}{3}} s_{13}e^{-i\delta}$.

Up to now, we have focused on deducing $\theta_{13} \neq 0$ via a
perturbation method starting from the basic tribimaximal neutrino
mixing matrix. Now, we attend the solar mass splitting. In our
framework of minimal perturbation matrix, we choose
$(M_\nu^P)_{12} = (M_\nu^P)_{21} = 0$. The first-order corrections
to the neutrino masses are obtained from $m^{(1)}_i\delta_{ij}=
~<\nu^{(0)}_i|M_{\nu}^P|\nu^{(0)}_j>$. We seek that the
first-order of neutrino mass corrections as
\begin{equation}
m^{(1)}_1 = m^{(1)}_3 = 0 ~~~{\rm and} ~~~m^{(1)}_2 \neq 0.
\label{efirstm}
\end{equation}

Therefore, in the mass basis, this implies that $(M_\nu^P)_{22}
\neq 0$ and other diagonal elements of the perturbation matrix to
be zero. Such a correction certifies to split  solar neutrino
masses, in which $ m^{(1)}_2=m_2 - m_1 $, and
 $\delta m^2 $ to be positive.

Consequently, in the mass basis, we have the final form of the
perturbation mass matrix, at point D, as follows
\begin{equation}\label{eM2}
{M_\nu^P}|_{\text{point
D}} =m_0~{s}_{13}\left(\begin{array}{ccc}0 & 0 & \sqrt{\frac{2}{3}}e^{-i\delta}\\
0 & \varepsilon &
-\sqrt{\frac{1}{3}}e^{-i\delta}\\\sqrt{\frac{2}{3}}e^{-i\delta} &
-\sqrt{\frac{1}{3}}e^{-i\delta} &0\end{array}\right),
\end{equation}
where $\varepsilon \equiv \frac{m^{(1)}_2}{m_0~{s}_{13}}$ is a
dimensionless parameter which relates the solar mass splitting,
$m^{(1)}_2$, to $s_{13}$. In the next section, we will utilize
$\varepsilon$ to compute the order of $s_{13}$.
In general $m^{(1)}_2$, the solar mass splitting, can be complex
as; $m^{(1)}_2 \equiv |m^{(1)}_2| ~\exp(i \varphi)$. Therefore, we
can write $m_2 = m^{(0)}_2 + m^{(1)}_2 \equiv |m_2| ~\exp(i \phi)$
and obtain neutrino mass spectrum at point D as follows,
\begin{eqnarray}
|m_1|
&=&|m^{(0)}_1|=m_0,~~~~~~~|m_3|=|m^{(0)}_3|=0,\nonumber\\|m_2| &=&
\left[(m^{(0)}_1)^2 + (|m^{(1)}_2|)^2 + 2 m^{(0)}_1 |m^{(1)}_2|
\cos\varphi\right]^{1/2}, \;\;  \phi = \tan^{-1}\left[{|m^{(1)}_2|
~\sin\varphi  \over m^{(0)}_1 + |m^{(1)}_2|  ~\cos
\varphi}\right]. \label{em2}
\end{eqnarray}

For the Dirac neutrinos, the phase $\phi$  can be removed but for
the Majorana neutrinos this phase remains as Majorana phase and
contribute to CP violation.


Now, let us rewrite $M_\nu^P$ [which is given by Eq. (\ref{eM2})
and it was calculated in the mass basis] in the flavor basis by
employing the relation between mass and flavor basis as

\begin{equation}\label{eM3} {M_\nu^{P^{(f)}}}|_{\text{point
D}} =m_0
~s_{13}\left[\frac{e^{-i\delta}}{\sqrt{2}}\left(\begin{array}{ccc}0 &1& -1\\
1& 0 & 0\\-1 &
0 &0\end{array}\right)+\frac{\varepsilon}{3}\left(\begin{array}{ccc}1 & 1 & 1\\
1 & 1 & 1\\1 & 1 &1\end{array}\right)\right],
\end{equation}
where the first and the second terms of $M_\nu^{P^{(f)}}$ are
responsible for 1-3 mixing angle and for solar neutrino mass
splitting, respectively. We observe that $M_\nu^{P^{(f)}}$ in Eq.
(\ref{eM3}) violates magic and $\mu-\tau$ symmetry of the total
neutrino mass matrix. Therefore, by using degenerate perturbation
theory \cite{Schiff}, to order linear in $s_{13}$, and $M_\nu^P$
in Eq. (\ref{eM2}) we obtain the neutrino mixing matrix with CP
violation parameters in the lepton sector at point D, as follows

\begin{equation}\label{eU}
{U}|_{\text{point
D}} =U_{TBM}+s_{13}e^{i\delta}\left(\begin{array}{ccc}0 & 0 & e^{-2i\delta}\\
-\sqrt{\frac{1}{3}} & \sqrt{\frac{1}{6}} & 0\\\sqrt{\frac{1}{3}} &
-\sqrt{\frac{1}{6}} &0\end{array}\right),
\end{equation}

$U$ is unitary up to order $s_{13}$.

In \cite{18}, with a different point of view, this same structure
for the neutrino mixing matrix has been discussed and the
consistency with the observed mixing angles noted.


A rephasing-invariant measure of CP violation in neutrino
oscillation is the universal parameter $J$ \cite{J} given in
Eq.\,(\ref{eJ}), where its form is independent of the choice of
the Dirac or Majorana neutrinos.

Using Eq.\,(\ref{eJ}) and Eq.\,(\ref{eU}) the expression for $J$
simplifies to,

\begin{eqnarray}\label{eJJ}\vspace{.5cm}
  {J}|_{\text{point
D}}&=&-{1 \over 3 \sqrt{2}}s_{13} \sin \delta,
\end{eqnarray}
as mentioned before, to have CP violation in the lepton sector,
both $s_{13}$ and $\delta$ must take nonzero values.

\section{Comparison with experimental data}\label{sec3}
In this section, we compare the results obtained from our
generalized FL model at remarkable point D (while $B=0$) with the
available experimental data in Eq.\,(\ref{exp}). It is important
to note that in the generalized FL model \cite{me1} almost there
are some numerical predictions for neutrino parameters while
$B\neq0$. Therefore, in this section, we compare our herein
results with the corresponding ones obtained in \cite{me1}, too.
Let us accomplish this in two steps.

In the first step, we compute the allowed ranges for the
parameters of the neutrino mass matrix along with the perturbation
term. We do this, by comparing and mapping our herein results of
neutrino masses with neutrino mass constraints obtained from the
available experimental data for $\delta m^{2}$ and $\Delta m^{2}$
in Eq.\,(\ref{exp}) respectively.

Therefore, the allowed ranges for $m_0$ and $g$ are obtained as,
\begin{eqnarray}\label{em0g}
m_0&\approx& \pm(4.87-5.03)10^{-2}eV,\nonumber\\
g&\approx&\pm(0.812-0.838)10^{-2}eV,
\end{eqnarray}
which agree well with the allowed ranges for $m_1$ and $m_0$ that
obtained in \cite{me1}. Also, we see that we can have $m_0>0$ and
$m_0<0$.

\begin{center}
\begin{figure}[th] \includegraphics[width=8cm]{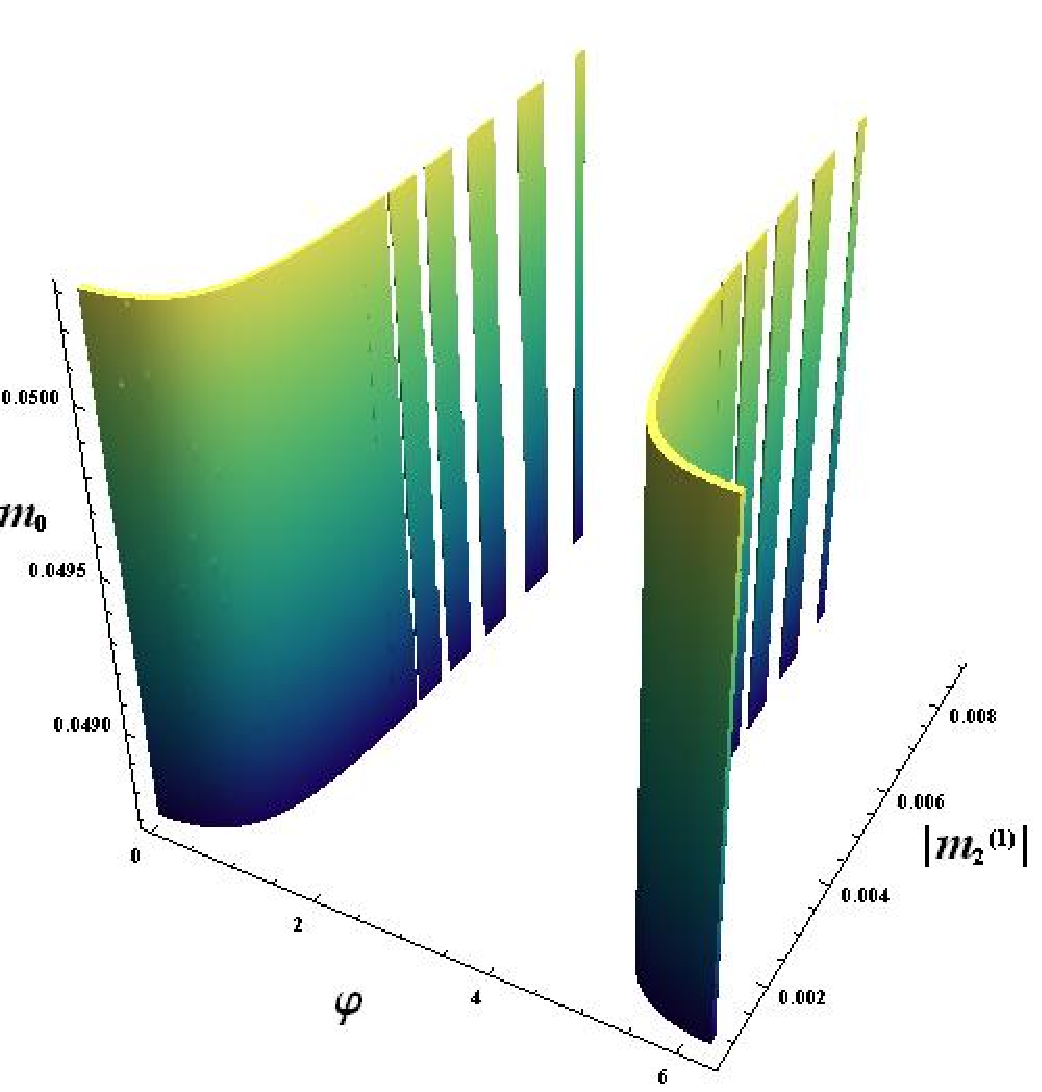}
\caption{\label{fig.2} \small
  (color online). In this figure, the whole region of the
$m^{(1)}_2-\varphi$ plane which is allowed by our model along with
the allowed range of $m_0>0$ is shown. Each color curve implies a
value of $m_0$ in the rang $(4.87-5.03)10^{-2}eV$ in the
${m^{(1)}_2}^2+2m_0m^{(1)}_2\cos\varphi$. When $m_0>0$, the
overlap region of the experimental values for $\Delta m_{21}^{2}$
with our model are two tiny regions. These regions are the
semi-symmetry of each other.}
  \label{geometry}
\end{figure}
\end{center}

\begin{center}
\begin{figure}[th] \includegraphics[width=8cm]{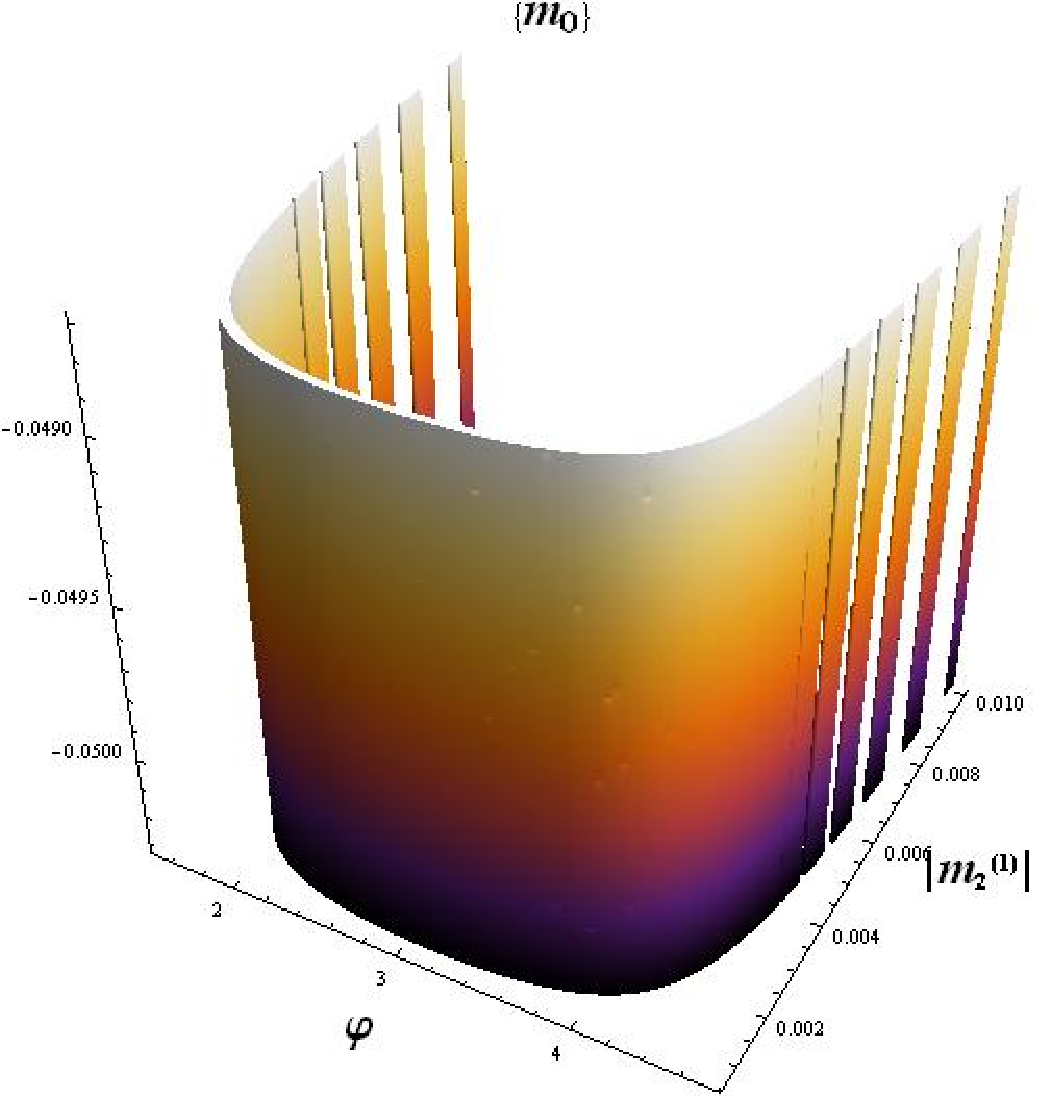}
\caption{\label{fig.3} \small
  (color online). In this figure, the whole region of the
$m^{(1)}_2-\varphi$ plane which is allowed by our model along with
the allowed range of $m_0<0$ is shown. Each color curve implies a
value of $m_0$ in the rang $-(4.87-5.03)10^{-2}eV$ in the
${m^{(1)}_2}^2+2m_0m^{(1)}_2\cos\varphi$. When $m_0<0$, the
overlap region of the experimental values for $\Delta m_{21}^{2}$
with our model is a tiny region.}
  \label{geometry}
\end{figure}
\end{center}

In Figure \ref{fig.2} we have shown the limits imposed by $\delta
m^{2}$ and $m_0>0$ on $m^{(1)}_2$ and in Figure \ref{fig.3} have
shown the limitations for $m_0<0$. In Figure \ref{fig.2} and
\ref{fig.3} we have plotted the overlap of $\delta m^{2}$ in
Eq.\,(\ref{em2}) by using the experimental data in
Eq.\,(\ref{exp}) with the allowed ranges of $m_0$ in
Eq.\,(\ref{em0g}) onto the $m^{(1)}_2$ and $\varphi$ perturbation
parameter space. Therefore, when $m_0>0$, our results for the
perturbation parameters are given by

\begin{equation}\label{empp}
m^{(1)}_2\approx (0.073-0.9)10^{-2}eV,
\end{equation}

and when $m_0<0$, our results are as follows

\begin{equation}\label{empq}
m^{(1)}_2\approx (0.073-0.9)10^{-2}eV.
\end{equation}

A point that should not be ignored about the allowed obtained
ranges of $m^{(1)}_2$ in Eq.\,(\ref{empp}) and Eq.\,(\ref{empp})
is that the obtained values of $m^{(1)}_2$ must predict the values
of $\sin \theta_{13}$, that to be consistent with the experimental
data in Eq.\,(\ref{exp}). To check this issue, we examine the
parameter $\varepsilon \equiv \frac{m^{(1)}_2}{m_0~{s}_{13}}$. We
assume that all nonzero components of the perturbation matrix in
Eq.\,(\ref{eM2}) should be in a similar order. We may then expect
$\varepsilon~\sim~{{\cal
 O}(1)}$, therefore by using the order of $m_0$ and $m^{(1)}_2$ from the
 previous equations we must obtain $\sin \theta_{13} \sim {\cal
 O}(10^{-1})$. We do this and find that the whole range of
 $m^{(1)}_2$ in Eq.\,(\ref{empp}) and Eq.\,(\ref{empp}) are not
 consistent with the experimental order of $\sin
 \theta_{13}$. Therefore, the consistent range of $m^{(1)}_2$ with the all of the experimental data for both cases, $m_0>0$ and $m_0<0$, is as
 follows,

\begin{eqnarray}\label{empp1}
|m^{(1)}_2|&\approx& (0.487-0.9)10^{-2}eV,~~~~~~\nonumber\\
\varphi&\approx&\{(83.12^{\circ}-91.72^{\circ})-(269.43^{\circ}-275.16^{\circ})\},~~~~~~\text{when}~~m_0>0\nonumber\\
~~~~~~~&\approx&\{(88.85^{\circ}-97.45^{\circ})-(263.69^{\circ}-269.43^{\circ})\},~~~~~~\text{when}~~m_0<0.
\end{eqnarray}

Therefore, we obtain $\sin \theta_{13} \sim {\cal
 O}(10^{-1})$ based on the values of $m_0$, $m^{(1)}_2$ (respectively in Eq.\,(\ref{em0g}) and Eq.\,(\ref{empp1})) and
considering $\varepsilon~\sim~{{\cal
 O}(1)}$. Further, employing Eq.\,(\ref{eU}),
and considering only the order of $\sin \theta_{13}$, we obtained
the allowed range of $\sin^2 \theta_{23}$ as follows,

\begin{equation}\label{es23}
\sin^2 \theta_{23}=
\frac{|U_{23}|^2}{1-|U_{13}|^2}\approx(0.505-0.549),
\end{equation}

which agrees well with the experimental data.

Therefore, not only we have obtained all the parameters of the
model at the remarkable point D, but also make predictions for
neutrino masses, see Eq.\,(\ref{emppp}). According to our
predictions in Eq.\,(\ref{emppp}), we find that both obtained
results related to $m_0>0$ and $m_0<0$ are in agreement with the
available experimental data, see Eq.\,(\ref{exp}).

\begin{eqnarray}\label{emppp}
m_1&=&m_0\approx \pm(4.87-5.03)10^{-2}eV,\nonumber\\
|m_2|&\approx&\{(4.95-5.08),(4.94-5.10)\}10^{-2}eV,\nonumber\\
~~~&\approx&\{(4.96-5.09),(4.95-5.12)\}10^{-2}eV,\nonumber\\
\phi&\approx&\{((5.60^{\circ})-(10.19^{\circ})),((-5.63^{\circ})-(-10.16^{\circ}))\},\nonumber\\
~~~&\approx&\{((-5.59^{\circ})-(-10.18^{\circ})),((5.61^{\circ})-(10.12^{\circ}))\},\nonumber\\
m_3&=&0,\nonumber\\
\delta m^{2}&\approx&\{(5.46-7.86),(6.87-7.10))\}10^{-5}eV^{2},\nonumber\\
~~~&\approx&\{(6.07-8.85),(7.86-9.13))\}10^{-5}eV^{2},\nonumber\\
|\Delta m^{2}|&\approx&(2.37-2.53)\times10^{-3}eV^{2}\footnote{In
this work, all the predictions agree well with the allowed ranges
of the neutrino masses that obtained in \cite{me1}.}.
\end{eqnarray}

Note that as mentioned before, $\phi$ is the origin of the
Majorana phases which is generated from the perturbation. We could
dispense with the overall phase, $e^{i\phi}$. Therefore, for
$m_0<0$ we would have two phases that emerge in the mass
eigenvalues in Eq.\,(\ref{emppp}) which are
$\rho=(\frac{\pi}{2}-\frac{\phi}{2})$ and
$\sigma=-\frac{\phi}{2}$. Whereas, for $m_0>0$
$\rho=\sigma=-\frac{\phi}{2}$. For the Dirac neutrinos, these
phases can be removed, and for the Majorana neutrinos, these
phases remain as Majorana phases and contribute to CP violation in
the lepton sector \footnote{All of our model predictions for Dirac
and Majorana neutrinos are the same, except Majorana phases in
general.} \cite{52}.

In the following, in the second step, by employing and referring
to the obtained values for the Jarlskog parameter which is
restricted by $0.027\leq J \leq 0.044$ in \cite{me1}, we find the
allowed ranges for $J$ and $\delta$. We do this by inserting the
experimental data of $\sin \theta_{13}$ in to Eq.\,(\ref{eJJ}) and
restricted $J$ in Eq.\,(\ref{eJJ}) to the allowed ranges of it in
\cite{me1}. Therefore, we obtain the allowed region of $J$ and
$\delta$ in our work as follows
\begin{eqnarray}\label{eJdelta}
J&\approx& (0.027-0.036),~~~~~~\nonumber\\
\delta&\approx&(229.30^{\circ}-312.42^{\circ}).
\end{eqnarray}

.

An important experimental result for the sum of the three light
neutrino masses has just been reported by the \textit{Planck}
measurements of the cosmic microwave background (CMB)
 at $95\%$ CL \cite{planck}, which is

 \begin{equation}\label{eplank}\vspace{.2cm}
\sum m_\nu<0.12eV \text{(Plank+WMAP+CMB+BAO)}.
\end{equation}
This sum in our model is

\begin{eqnarray}\label{emsum}
\sum
m_\nu&\approx&\{(0.0982-0.1011)eV-(0.0981-0.1013)eV\},~~~~~~\text{when}~~m_0>0\nonumber\\
~~~~~~~&\approx&\{(0.0983-0.1012)eV-(0.0982-0.1015)eV\},~~~~~~\text{when}~~m_0<0.
\end{eqnarray}
, which is exactly consistent with the constraint of the above.

\section{Conclusions}

In this paper we have recommenced a generalization of
Friedberg-Lee neutrino mass model, in which CP violation is
viable. Therefore, the elements of neutrino mass matrix could be
complex. Of course with the constraint that the mass eigenvalues
be real. We obtain and indicate the region in our parameter space
where CP violation is possible. It is worth mentioning, we have
studied and focused only at a substantial point (point D) on the
border of CP violation region at $\alpha=\beta=-\frac{1}{3}$ (see
figure (\ref{fig.1})). The features of point D are: (i) at point
D, we have a neutrino mass matrix with $S_3$ permutation family
symmetry, which is called Democratic matrix, in the parameter
space. (ii) The corresponding mixing matrix at point D is
$U_{TBM}$. (iii) At this point, due to the degenerate mass
spectrum, we have $\Delta m^2=0$ which is experimentally
unfavored.

Therefore, we improve our model to obtain the experimentally
favored neutrino mass spectrum by adding the breaking mass term
which preserve the twisted FL symmetry to the mass matrix at point
D. Hence, we obtain a neutrino mass spectrum without degeneracy
and the corresponding mixing matrix is $U_{TBM}$, also our method
predicts the massless third generation neutrino and inverted mass
hierarchy. In the following, our method is based on the $U_{TBM}$
in which the mixing angles (except $\theta_{13}$) is consistent
with the experimental data. The tribimaximal mixing matrix led us
to generate a neutrino mass matrix, ${{M_0}^\nu}|_{point D}$,
constrained by the elements of $U_{TBM}$ and the available
experimental data.

Therefore, ${{M_0}^\nu}$ is unperturbed mass matrix and loses the
solar neutrino mass splitting whereas it preserves magic and
$\mu-\tau$ symmetry features. In the following, by employing
perturbation method, we produce a perturbation matrix, at point D,
which breaks both the magic and $\mu-\tau$ symmetry, therefore CP
violation is happen. Our investigation proceeded as follows, we
obtained the complex elements of the perturbation mass matrix in
both the mass and flavor bases. In addition, We regenerate a
realistic neutrino mixing matrix, at point D, with non-zero the
Dirac phase. However, in this case, the $\mu-\tau$ symmetry is
softly broken, but still we have $\theta_{23}=45^\circ$.

For the purpose of getting our predictions regarding neutrino
masses at point D, we compared the results of our phenomenological
model with the available experimental data. We have display that
how our phenomenological model at point D whether or not is
compatible with experimental data. In our work, the most
restricting experimental data comes from the value of
$m_1=\sqrt{|\Delta m^2|}$ and the order of $\sin\theta_{13}$.
Mapping the allowed range of the experimental data of $\delta m^2$
onto the allowed region of our parameter space can designate valid
values for our parameters. Also, we have shown that only by
applying the order of $\sin\theta_{13}$ can restrict the allowed
region of the predicted solar mass splitting. Afterward, we can
predict the values of three masses, $m
_1\approx\pm(4.87-5.03)10^{-2}eV$,
$|m_2|\approx\{\{(4.95-5.08),(4.94-5.10)\}10^{-2}eV\}\cup\{\{(4.96-5.09),(4.95-5.12)\}10^{-2}eV\}$,
and $m_3=0$. The compatibility of the allowed ranges of our
parameters with the current experimental data shows that our model
has inverted hierarchy as, $m_3=0$. We also obtain predictions for
the CP violation parameters $\delta$, and $J$ (at point D). These
are $\delta\approx(229.30^{\circ}-312.42^{\circ})$, and $J\approx
(0.027-0.036)$. Our predictions are agree with the observational
data reported by Planck(+WMAP+CMB+BAO)experiment.

In our method, at an unusual point D, the mass matrix was saved
from having degenerate unfavored excremental eigenvalues and the
least possible perturbation matrix was obtained by employing a
fundamental process, and not only added by hand. Our predictions
for the neutrino masses and CP violation parameters could be
assayed in future neutrino experiments.

\section{Appendix A :Twisted Friedberg-Lee symmetry}
In this appendix, we give a detailed discussion how to obtain the
neutrino mass matrix used in the main part from the twisted FL
symmetry for Dirac neutrinos\cite{TFL}.

Let us consider the Dirac neutrino case
\begin{eqnarray}
 -{\cal L}_D=\bar{\nu}_{Li} M^D_{ij} \nu_{Rj} + h.c.\ .
\end{eqnarray}
For Dirac neutrinos, in general, the twisted FL symmetry can be
imposed on the left- and right-handed neutrinos separately as
\begin{eqnarray}
 &&\nu_{Li} \rightarrow \nu_{Li}^{'}=S_{ij}^L\nu_{Lj}+\Lambda_{Lj}z\ ,\\
 &&\nu_{Ri} \rightarrow \nu_{Ri}^{'}=S_{ij}^R\nu_{Rj}+\Lambda_{Rj}z\ .
\end{eqnarray}
Two independent $\mu - \tau$ permutation symmetries make the Dirac
mass matrix as
\begin{eqnarray}
 M^D=
 \left(\begin{array}{ccc}
 D & -2C & -2C \\
 -2B & -A & -A \\
 -2B & -A & -A
 \end{array}\right)\ ,
\end{eqnarray}
while the translational symmetries lead to the conditions
\begin{eqnarray}
 &&M^D_{ij}\ \Lambda_{Rj}=
 \left(\begin{array}{ccc}
 D & -2C & -2C \\
 -2B & -A & -A \\
 -2B & -A & -A
 \end{array}\right)
 \left(\begin{array}{c}
 \Lambda_{R1} \\ \Lambda_{R2} \\ \Lambda_{R3}
 \end{array}\right)=0\ ,
\end{eqnarray}
and\begin{eqnarray}
 &&\Lambda_{Li}\ M^D_{ij}=
 \left(\begin{array}{ccc}
 \Lambda_{L1} & \Lambda_{L2} & \Lambda_{L3}
 \end{array}\right)
 \left(\begin{array}{ccc}
 D & -2C & -2C \\
 -2B & -A & -A \\
 -2B & -A & -A
 \end{array}\right)=0\ .
\end{eqnarray}
The resulting form of the mass matrix depends on the correlations
among $\Lambda_{Li}$ and $\Lambda_{Ri}$ again. In this letter, we
have assumed the uniform translation, that is
$\Lambda_{Li}\propto(1,1,1)$ and $\Lambda_{Ri}\propto(1,1,1)$.
Then, the mass matrix of the Dirac neutrino takes the form
\begin{eqnarray}
  M^D=C
 \left(\begin{array}{ccc}
 4 & -2 & -2 \\
 -2 & 1 & 1 \\
 -2 & 1 & 1
 \end{array}\right)\ .
\end{eqnarray}



\begin{thebibliography}{99}
\bibitem{exp}
SNO Collaboration, Q.R. Ahmad, et al., Phys. Rev. Lett. 89 (2002)
011301; For a review, see: C.K. Jung, et al., Annu. Rev. Nucl.
Part. Sci. 51 (2001) 451; KamLAND Collaboration, K. Eguchi, et
al., Phys. Rev. Lett. 90 (2003) 021802; K2K Collaboration, M.H.
Ahn, et al., Phys. Rev. Lett. 90 (2003) 041801.
\bibitem{exp1}
D. A. Dwyer [Daya Bay Collaboration], Nucl. Phys. Proc. Suppl.
235-236, 30 (2013) [arXiv:1303.3863 [hep-ex]]. F. P. An et al.
[DAYA-BAY Collaboration], Phys. Rev. Lett. 108, 171803 (2012)
[arXiv:1203.1669 [hep-ex]]; J. K. Ahn et al. [RENO Collaboration],
Phys. Rev. Lett. 108, 191802 (2012) [arXiv:1204.0626[hep-ex]].
\bibitem{mixing}
J. Schechter and J. W. F. Valle Phys.Rev. D22 (1980) 2227; H.
Fritzsch and Z.Z. Xing, Phys. Lett. B 517, 363 (2001). Particle
Data Group, W.M. Yao et al., J. Phys. G 33, 1 (2006).
\bibitem{FL}
R. Friedberg, T.D. Lee, High Energy Phys. Nucl. Phys. 30 (2006)
591; arXiv:hep-ph/0606071.
\bibitem{TBM}
P. F. Harrison at al., Phys. Lett. B458, 79 (1999); P. F. Harrison
et al., Phys. Lett. B530, 167 (2002); Z-z. Xing, Phys. Lett. B533,
85 (2002); P. F. Harrison and W. G. Scott, Phys. Lett. B535, 163
(2002); P. F. Harrison and W. G. Scott, Phys. Lett. B557, 76
(2003); X.-G. He and A. Zee, Phys. Lett. B560, 87 (2003).
\bibitem{FL2}
C. S. Huang, T. Li, W. Liao and S. H. Zhu, Phys. Rev. D 78, 013005
(2008).



\bibitem{demo1}
 H. Harari, H. Haut and J. Weyers,
Phys. Lett. B78 (1978) 459; Y. Koide, Phys. Rev. D28 (1983) 252;
ibid. D39 (1989) 1391; P. Kaus and S. Meshkov, Mod. Phys. Lett. A3
(1988) 1251 [Erratum-ibid. A4 (1989) 603]; L. Lavoura, Phys. Lett.
B228 (1989) 245; M. Tanimoto, Phys.  Rev. D41 (1990) 1586; G.C.
Branco, J.I. Silva- Marcos and M.N. Rebelo, Phys. Lett. B237
(1990) 446; H. Fritzsch and J. Plankl, Phys. Lett. B237 (1990)
451; H. Fritzsch and Z.Z. Xing, Phys. Lett. B372 (1996) 265; M.
Fukugita, M. Tanimoto and T. Yanagida, Phys.  Rev. D57 (1998)
4429; R.N. Mohapatra and S. Nussinov, Phys.  Lett. B441 (1998)
299;  M. Tanimoto, Phys. Lett. B 483 , 417 (2000); P.F. Harrison
and W.G. Scott, Phys.  Lett. B557 (2003) 76; F. Caravaglios and S.
Morisi, arXiv:hep-ph/0503234; N. Haba, K. Yoshioka, Nucl.Phys.
B739 (2006), arXiv:hep-ph/0511108v2.
\bibitem{8}
See, for example, P. Ciafaloni, M. Picariello, A. Urbano and E.
Torrente-Lujan, Phys. Rev. D 81 , 016004 (2010) [arXiv:0909.2553
[hep-ph]]; P. H. Frampton , T. W. Kephart and S. Mat- suzaki,
Phys. Rev. D 78 , 073004 (2008) [arXiv:0807.4713 [hep-ph]]; F.
Plentinger , G. Seidl and W. Winter, JHEP 0804 , 077 (2008)
[arXiv:0802.1718 [hep-ph]]; F. Bazzocchi, S. M orisi and M.
Picariello, Phys. Lett. B 659 , 628 (2008) [arXiv:0710.2928
[hep-ph]]; G. Altarelli, F. F er- uglio, Nucl.Phys. B 720 , 64
(2005); G. Altarelli and F. Feruglio, Nucl. Phys. B 741 , 215
(2006) [hep-ph/0512103].
\bibitem{ttbm}
 G. Altarelli and F. Feruglio, Nucl. Phys. B
741 , 215 (2006) [hep-ph/0512103];  B. Adhikary, B. Brahmachari,
A. Ghosal, E. Ma and M. K. Pa rida, Phys. Lett. B 638 , 345 (2006)
[hep-ph/0603059]; F. Bazzocchi, S. Morisi and M. Picariello, Phys.
Lett. B 659 , 628 (2008) [arXiv:0710.2928 [hep-ph]]; S. F. King,
Phys. Lett. B 659 , 244 (2008) [arXiv:0710.0530 [hep-ph]]. S.
Pakvasa, W. Rod ejo- hann, T. Weiler, Phys. Rev. Lett. 100 ,
111801 (2008);  E. Ma, Phys. Lett. B 660 , 505 (2008)
[arXiv:0709.0507 [hep-ph]];  F. Plentinger, G. Seidl and W.
Winter, JHEP 0804 , 077 (2008) [arXiv:0802.1718 [hep-ph]]; N.
Haba, R. Takahashi, M. Tanimoto and K. Yoshioka, Phys. Rev . D 78
, 113002 (2008) [arXiv:0804.4055 [hep-ph]]; C. H. Albright, W.
Rodejohann Eur. Phys. J. C 62 , 599-608 (2009); S. Boudjemaa and
S. F. King, Phys. Rev. D 79 , 033001 (2009) [arXiv:0808.2782
[hep-ph]]; S. Goswami, S. T. Petcov, S. Ray and W. Rodejohann,
Phys. Rev. D 80 , 053013 (2009) [arXiv:0907.2869 [hep-ph]]; C. H.
Albright, A. Dueck, W. Rodejohann, Eur. Phys. J. C 70 , 1099-1110
(2010); E. Ma and D. Wegman, Phys. Rev. Lett. 107 , 061803 (2011)
[arXiv:1106.4269 [hep-ph]]; X. He, A. Zee, Phys. Rev. D 84 ,
053004 (2011); ]. D. Meloni, F. Plentinger and W . Winter, Phys.
Lett. B 699 , 354 (2011) [arXiv:1012.1618 [hep-ph]]; D. Marzocca,
S. T. Petc ov, A. Romanino and M. Spinrath, JHEP 1111 , 009 (2011)
[arXiv:1108.0614 [hep-ph]]; S. Gupta, A. S. Joshipura and K. M.
Patel, Phys. Rev. D 85 , 031903 (2012) [arXiv:1112.6113 [hep-ph]];
S. Dev, R. R. Gautam and L. Singh, Phys. Lett. B 708 , 284 (2012)
[arXiv:1201.3755 [hep-ph]]; ]; S. -F. Ge, D. A. Dicus and W. W. R
epko, Phys. Rev. Lett. 108 , 041801 (2012) [arXiv:1108.0964
[hep-ph]]; T. Araki and Y. F . Li, Phys. Rev. D 85 , 065016 (2012)
[arXiv:1112.5819 [hep-ph]]; G. C. Branco, R. G. Felipe, F. R.
Joaquim and H. Serodio, arXiv :1203.2646 [hep-ph]; B. Grinstein
and M. Trott, arXiv:1203.4410 [hep-ph]; Gayatri Ghosh, Nuclear
Physics B(2022) [2106.12503 [hep-ph]]; Ph. Wilina, M. Shubhakanta
Singh, N. Nimai Singh, Deviations from Tribimaximal and Golden
Ratio mixings under radiative corrections of neutrino masses and
mixings (2022) [2205.01936 [hep-ph]];E. Barradas-Guevara, O.
Félix-Beltrán, F. Gonzalez-Canales, Deviation to the
Tri-Bi-Maximal flavor pattern and equivalent classes (2022)
[2204.03664 [hep-ph]]; Hui-Chao Bao et al, 2022 Commun. Theor.
Phys. 74 055201.
\bibitem{J}
C. Jarlskog, Phys. Rev. Lett. 55 (1985) 1039.
\bibitem{theoretical}
Z. Z. Xing, H. Zhang, S. Zhou, Phys Lett B 641 (2006) 189; T.
Baba, M. Yasue, Phys. Rev. D75, 055001 (2007); Z. Z. Xing, H.
Zhang, S. Zhou, Int. J. Mod. Phys. A 23, 3384(2008).
\bibitem{phase1}
J. Schechter and J.W. F. Valle, Phys. Rev. D 22, 2227 (1980);
Physics of neutrinos and applications to as- trophysics, Masataka
Fukugita and Tsutomu Yanagida, Springer, (2003); S. Nasri, J.
Schechter and S. Moussa , Phys. Rev. D 70, 053005(2004).
\bibitem{me1}
 N. Razzaghi, S. S. Gousheh, Phys. Rev. D 86,
(2012) 053006.
\bibitem{TFL}
Takeshi Araki, Ryo Takahashi, Eur. Phys. J. C63:521-526,(2009).
\bibitem{meuni}
Razzaghi, N.; Rasouli, S.M.M.; Parada, P.; Moniz, P. Generating CP
Violation from a Modified Fridberg-Lee Model. Universe 2022, 8,
448.
\bibitem{Schiff} L. I. Schiff, {\em Quantum Mechanics} (Third ed.),
McGraw-Hill (1968).
\bibitem{18}
Xing, Z.-Z. A Shift from Democratic to Tri-bimaximal Neutrino
Mixing with Relatively Large $\theta_{13}$. Phys. Lett. B 2011,
696, 232-236.
\bibitem{52}
Fukugita, M.; Yanagida, T. Physics of Neutrinos and Applications
to Astrophysics; Springer: New York, NY, USA, 2003.
\bibitem{planck}
N. Aghanim et al. [Planck], Astron. Astrophys. 641, A6 (2020),
[arXiv:1807.06209 [astro-ph.CO]].
















\end{thebibliography}
\end{document}